\documentclass[12pt]{article}
\usepackage{epsfig}
\usepackage{amsmath}
\usepackage{amssymb}
\begin{document}
\begin{center}
\LARGE
\textbf{Does Quantum Mechanics Need
Interpretation?}\\[1cm]
\large
\textbf{Louis Marchildon}\\[0.5cm]
\normalsize
D\'{e}partement de physique,
Universit\'{e} du Qu\'{e}bec,\\
Trois-Rivi\`{e}res, Qc.\ Canada G9A 5H7\\
email: marchild$\hspace{0.3em}a\hspace{-0.8em}
\bigcirc$uqtr.ca\\
\end{center}
\bigskip
\begin{abstract}
Since the beginning, quantum mechanics
has raised major foundational and interpretative problems.
Foundational research has been an important factor in the development
of quantum cryptography, quantum information theory and, perhaps
one day, practical quantum computers.  Many believe that, in turn,
quantum information theory has bearing on foundational research.
This is largely related to the so-called epistemic view of quantum
states, which maintains that the state vector represents information
on a system and has led to the suggestion that quantum theory needs no
interpretation.  I will argue that this and related approaches fail
to take into consideration two different explanatory functions
of quantum mechanics, namely that of accounting for classically
unexplainable correlations between classical phenomena and that
of explaining the microscopic structure of classical objects.
If interpreting quantum mechanics means answering the question,
``How can the world be for quantum mechanics to be true?'',
there seems to be no way around it.
\end{abstract}
\section{Introduction}
Ever since it was proposed more than 80 years ago,
quantum mechanics has raised great challenges both
in foundations and in applications.  The latter have
been developed at a very rapid pace, opening up
new vistas in most branches of physics as well as in
much of chemistry and engineering.  Substantial progress
and important discoveries have also been made in foundations,
though at a much slower rate.  The measurement
problem, long-distance correlations, and the meaning
of the state vector are three of the foundational
problems on which there has been and still is lively debate.

It is fair to say that foundational studies have
largely contributed to the burgeoning of quantum
information theory, one of the most active areas of
development of quantum mechanics in the past 25 years.
Quantum information is dependent on entanglement,
whose significance was brought to light through the
Einstein-Podolsky-Rosen argument~\cite{einstein}.  The
realization that transfer protocols based
on quantum entanglement may be absolutely
secure has opened new windows in the field of
cryptography~\cite{bennett}.  And the development
of quantum algorithms thought to be exponentially
faster than their best classical counterparts has
drawn great interest in the construction of quantum
computers~\cite{shor}.  These face up extraordinary
challenges on the experimental side.
But attempts to build them are likely to throw
much light on the fundamental process of
decoherence and perhaps on the limits of quantum mechanics
itself~\cite{hooft,leggett}.

Along with quantum information theory came also a reemphasis
of the view that the wave function (or state vector, or
density matrix) properly represents knowledge, or
information~\cite{rovelli,fuchs1,fuchs2}.
This is often called the \emph{epistemic view} of quantum states.
On what the wave function is knowledge of, proponents of
the epistemic view do not necessarily agree.
The variant most relevant to the present discussion is
that rather than referring to objective
properties of microscopic objects (such as
electrons, photons, etc.), the wave function
encapsulates probabilities of results of eventual macroscopic
measurements.  The Hilbert space formalism of quantum
mechanics is taken as complete, and its objects in
no need of a realistic interpretation.  Additional
constructs, like value assignments~\cite{vermaas},
multiple worlds~\cite{everett}, or Bohmian
trajectories~\cite{bohm} are viewed
as superfluous at best.

Just like foundational studies have contributed to
the development of quantum information theory,
many investigators think
that the latter can help in solving
the foundational and interpretative problems of
quantum mechanics.  A number of proponents of the epistemic
view believe that it considerably attenuates, or even completely
solves, the problems of quantum measurement, of
long-distance correlations, and of the meaning of the state
vector.  Their arguments will be briefly summarized in
Sec.~2.  I will then argue, in Sec.~3, that the epistemic
view and related approaches fail to take into consideration
that quantum mechanics has two very different explanatory
functions, that of accounting for classically unexplainable
correlations between classical phenomena, and that
of explaining the microscopic structure of classical objects.
In Sec.~4, I will ask the question of what it means to
interpret quantum mechanics, or any scientific theory for that
matter.  Drawing from the so-called semantic view of theories,
I will argue than interpreting quantum mechanics means
answering the question, ``How can the world be for quantum mechanics
to be true?''  I will conclude that so construed, an interpretation
can hardly be dispensed with.\footnote{This paper reformulates
the arguments made in~\cite{marchildon1}, \cite{marchildon2},
and~\cite{marchildon3}, where additional material can be found.}
\section{The epistemic and related views}
Let us first examine the arguments that
advocates of the epistemic view offer to solve
the foundational and interpretative problems of
quantum mechanics.  I should point out that they
do not all attribute the same strength and generality to
these arguments.  Some advocates believe that the
problems are completely solved by the epistemic view,
while others are of the opinion
that they are just attenuated.  This distinction,
however, is not crucial to our purpose, and I will simply
give the arguments as they are typically formulated.

The first problem that is addressed by the epistemic
view is the one of the interpretation of the state
vector (or state operator, or wave function).
As the name suggests, the state vector is normally
interpreted as representing the state of quantum
systems.  It is a matter of debate whether the state it
represents pertains to an individual system or to statistical
ensembles of systems~\cite{ballentine}.  But the epistemic
view, which goes back at least to writings of
Heisenberg~\cite{heisenberg}, claims that it represents
neither.  It denies that the (in this context utterly
misnamed) state vector represents the state of a
microscopic system.  Rather, it represents
knowledge about the probabilities of results of
measurements performed in a given context with a
macroscopic apparatus, in other words, information about
``the potential consequences of our experimental
interventions into nature''~\cite{fuchs2}.  This
is often set in the framework of a Bayesian approach,
where probability is interpreted in a subjective way.

The epistemic view also addresses the notorious
measurement problem.  Broadly speaking, the problem
is the following.  Suppose we want to describe, in a
completely quantum-mechanical way, the process of
measuring a physical quantity $Q$ pertaining to a
microscopic system.  For simplicity, assume that the
spectrum of $Q$ is discrete and nondegenerate, and
that the normalized eigenvector $|q_i\rangle$
corresponds to the eigenvalue $q_i$.  The measurement
apparatus should then also be considered as a quantum
system, which comes to interact with the microscopic system.
Let $|\alpha_0 \rangle$ denote the initial state of the
apparatus.  The interaction will represent a faithful
measurement of $Q$ if the combined system evolves like
\begin{equation}
|q_i \rangle |\alpha_0 \rangle \rightarrow
|q_i \rangle |\alpha_i \rangle ,
\end{equation}
where $|\alpha_i \rangle$ represents a state of the
apparatus wherein the pointer shows the value
$\alpha_i$ (with $\alpha_i \neq \alpha_j$ 
if $i \neq j$).

If the Schr\"{o}dinger equation is universally
valid, the combined evolution of the microscopic
system and macroscopic apparatus is unitary (assuming,
unrealistically, that they form together a closed system).
But then, an initial state involving the superposition
of several eigenstates of an observable of the microscopic system
evolves into a final state involving a superposition of
macroscopically distinct states of the apparatus
(or of the apparatus and environment in more
realistic situations).  Explicitly,
\begin{equation}
\left\{ \sum_i c_i |q_i \rangle \right\}
|\alpha_0 \rangle \rightarrow
\sum_i c_i |q_i \rangle |\alpha_i \rangle .
\end{equation}
One solution to this problem appeals to the collapse
of the state vector~\cite{neumann}, in which the
Schr\"{o}dinger equation breaks down and only one term
of the macroscopic superposition (e.g.\
$|q_j \rangle |\alpha_j \rangle$) remains.

How does the epistemic view deal with the
measurement problem?  It does so by construing
the collapse of the state vector not as a
physical process, but as a change of
knowledge~\cite{peierls}.  Insofar as the state
vector is interpreted as objectively describing
the state of a physical system, its abrupt change
in a measurement
implies a similar change in the system, which
calls for explanation.  If, on the other hand,
and in line with a Bayesian view,
the state vector describes knowledge of
conditional probabilities (i.e.\ probabilities of
future macroscopic events conditional on past
macroscopic events), then as long as what is 
conditionalized upon remains the same, the state
vector evolves unitarily.  It collapses when the
knowledge base changes (this is Bayesian updating),
thereby simply reflecting
the change in the conditions being held fixed
in the specification of probabilities.

A third problem which is addressed by the epistemic view
is the one of long-distance correlations~\cite{fuchs2,bloch}.
Consider the realization of the Einstein-Podolsky-Rosen
setup in terms of two spin 1/2 particles (labelled $A$
and $B$), where the state vector $|\chi\rangle$
of the compound system is an eigenstate of the
total spin operator with eigenvalue zero.  In this case
\begin{equation}
|\chi\rangle = \frac{1}{\sqrt{2}}
\left\{ |+; \mathbf{n} \rangle \otimes |-; \mathbf{n} \rangle -
|-; \mathbf{n} \rangle \otimes |+; \mathbf{n} \rangle \right\} .
\end{equation}
Here the first vector in a tensor product refers to
particle $A$ and the second vector to particle $B$.
The vector $|+; \mathbf{n} \rangle $, for instance,
stands for an eigenvector of the $\mathbf{n}$-component
of the particle's spin operator, with eigenvalue
$+1$ (in units of $\hbar/2$).
The unit vector $\mathbf{n}$ can point in any direction,
a freedom which corresponds to the
rotational symmetry of~$|\chi\rangle$.

Suppose Alice measures the $\mathbf{n}$-component
of $A$'s spin and obtains the value $+1$.  Then
she can predict with certainty that if Bob measures
the same component of $B$'s spin, he will obtain the
value $-1$.  If the state vector represents the
objective state of a quantum system, it then seems that
$B$'s state changes immediately upon Alice's obtaining
her result, and this no matter how far apart $A$ and
$B$ are.  Since the word ``immediately'', when
referring to spatially separated events, is not a
relativistically invariant concept, such a mechanism is
not easy to reconcile with the theory of special relativity.

In the epistemic view, what changes when Alice
performs a measurement is Alice's knowledge.
Bob's knowledge will change either if he himself
performs a measurement, or if Alice sends him
the result of her measurement by conventional
means.  Hence no information is transmitted
instantaneously, and there is no physical collapse
on an equal time or spacelike hypersurface.

Related to the epistemic view is the idea of
\emph{genuine fortuitousness}~\cite{ulfbeck,bohr},
a radically instrumentalist view of quantum
mechanics.  The idea ``implies
that the basic event, a click in a counter, comes
without any cause and thus as a discontinuity in 
spacetime''~\cite[p.~405]{bohr}.  Indeed
\begin{quote}
[i]t is a hallmark of the theory based on genuine
fortuitousness that it does not admit physical variables.
It is, therefore, of a novel kind that does not deal
with things (objects in space), or measurements, and
may be referred to as the theory of no things. (p.~410)
\end{quote}

Such approaches to the interpretation of quantum mechanics
are to be contrasted with realist views like Bohmian
mechanics~\cite{bohm}.  Here particles are taken to
exist and they follow deterministic trajectories.
For illustration, consider a set of $N$ nonrelativistic
spin-zero particles interacting through a potential.
The system's wave function is a solution of the
Schr\"{o}dinger equation.  We can write it in polar
form as
\begin{equation}
\Psi (\mathbf{r}_1, \ldots, \mathbf{r}_N, t)
= \rho \exp (i S /\hbar) .
\end{equation}
The equation of motion of particle $i$ is given by
\begin{equation}
m_i \mathbf{v}_i = \boldsymbol{\nabla}_i S .
\label{motion}
\end{equation}
In general, $S$ is a nonadditive and nonseparable
function of all particle coordinates $\mathbf{r}_i$.
This can be shown to account for nonlocal effects
like long-distance correlations.

In Bohmian mechanics, the statistical properties
of quantum mechanics arise from an incomplete
knowledge of the system's initial conditions.  Indeed
the particles' initial positions, although well-defined,
are ``hidden'' to any observer.  What is known is their
statistical distribution, taken to be proportional
to the absolute square of the wave function.  It
can be shown that Bohmian mechanics exactly
reproduces the statistical results of quantum
mechanics.  In particular~\cite{philippidis},
interference fringes in Young's two-slit setup follow
directly from~(\ref{motion}).

If Bohmian mechanics exactly reproduces the statistical
results of quantum mechanics, aren't the trajectories
superfluous, and shouldn't they be discarded?
The analogy has been made between such trajectories
and the concept of the ether prevalent at the turn
of the twentieth century~\cite{bub1,bub2}.
H.~A.~Lorentz and his contemporaries
viewed electromagnetic phenomena as taking place
in a hypothetical medium called the ether.  From
this, Lorentz developed a description of
electromagnetism in moving reference frames,
and he found that the motion is
undetectable.  Following Einstein's formulation of
the electrodynamics of moving
bodies, the ether was recognized as
playing no role, and was henceforth discarded.
So should it be, according to most proponents of the
epistemic view of quantum states, with interpretations
of quantum mechanics that posit observer-independent
elements of reality like Bohmian trajectories.
They predict no empirical differences
with the Hilbert space formalism, and therefore
should be discarded.
\section{Two explanatory functions}
To examine how appropriate are the epistemic
and related views of quantum mechanics, it is important
to properly understand the explanatory role of
quantum mechanics as a physical theory.
Although all measurements are made
by means of macroscopic apparatus, quantum
mechanics is used, as an explanatory theory,
in two different ways: it is meant to explain
(i) nonclassical correlations between macroscopic
objects and (ii) the small-scale structure of
macroscopic objects.  That these two functions
are distinct is best shown by contrasting the world
in which we live with a hypothetical, closely
related one.

Roughly speaking, the hypothetical world is defined
so that (a) for all practical purposes,
all macroscopic experiments give results
that coincide with what we find in the real world,
and (b) its microscopic structure, if applicable,
is different from the one of the real world.  Let
us spell this out in more detail.

In the hypothetical world large scale objects,
i.e.\ objects much larger than atomic sizes, behave
just like large scale objects in the
real world.  The trajectories of baseballs and
airplanes can be computed accurately by means
of classical mechanics with the use of a
uniform downward force, air friction, and an
appropriate propelling force.  Waveguides and
antennas obey Maxwell's equations.  Steam engines and
heat pumps work according to the laws of
classical thermodynamics.  The motion of planets,
comets, and asteroids is well described by
Newton's laws of gravitation and of motion,
slightly corrected by the equations of
general relativity.

Close to atomic scales, however, these laws
may no longer hold.  Except for one restriction
soon to be spelled out, I shall not be specific
about the changes that macroscopic laws may or
may not undergo in the microscopic realm.  Matter, for
instance, could either be continuous down to
the smallest scales, or made of a small
number of constituent particles like our
atoms.  The laws of particles and fields
could be the same at all scales, or else
they could undergo significant changes
as we probed smaller and smaller distances.

\begin{sloppypar}
In the hypothetical world one can perform
experiments with pieces of equipment like 
Young's two-slit setup, Stern-Gerlach devices, or
Mach-Zehnder interferometers.  Let us focus on the
Young type experiment.  It makes use of two macroscopic
objects which we label $E$ and $D$.  These symbols could
stand for ``emittor'' and ``detector'' if it were not
that, as we shall see, they may not emit or detect
anything.  At any rate, $E$ and $D$
both have on and off states and work in the
following way.  Whenever $D$ is suitably
oriented with respect to $E$ (say, roughly
along the $x$ axis) and both are in
the on state, $D$ clicks in a more or less
random way.  The average time interval
between clicks depends on the distance $r$
between $D$ and $E$, and falls roughly as $1/r^2$.
The clicking stops if, as shown in Fig.~1, a shield of a
suitable material is placed perpendicularly
to the $x$ axis, between $D$ and $E$.
\end{sloppypar}

\begin{figure}[bt]
\begin{center}
\epsfig{file=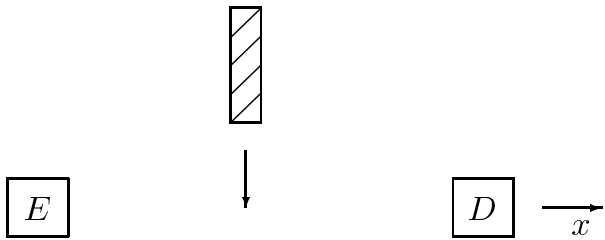,height=32mm,width=80mm}
\end{center}
\caption{Shielding material prevents $D$ from clicking}
\label{fig1}
\end{figure}

If holes are pierced through the shield, however,
the clicking resumes.  In particular, with
two small holes of appropriate size and
separation, differences in the clicking rate
are observed for small transverse displacements
of $D$ behind the shield.  A
plot of the clicking rate against $D$'s
transverse coordinate displays maxima and
minima just as in a wave interference pattern.
No such maxima and minima are observed, however,
if just one hole is open or if both holes
are open alternately.

At this stage everything happens as if $E$
emitted some kind of particles and $D$
detected them, and the particles behaved
according to the rules of quantum mechanics.
Nevertheless, we shall nor commit ourselves
to the existence or nonexistence of these
particles, except on one count.  Such particles,
if they exist, are not in any way related to
hypothetical constituents of the material
making up $D$, $E$, or the shield, or of any
macroscopic object whatsoever.  Whatever
the microscopic structure of macroscopic
objects is, it has nothing to do with what is
responsible for the correlations between
$D$ and $E$.

In a similar way, we can perform in the
hypothetical world experiments with Stern-Gerlach
devices, Mach-Zehnder interferometers, or other
setups used in the typical quantum-mechanical
investigations carried out in the real world.
Correlations are observed between initial states
of ``emittors'' and final states of ``detectors''
which are unexplainable by classical mechanics
but follow the rules of quantum mechanics.  We
assume again that, if these correlations have
something to do with the emission and absorption
of particles, these are in no way related
to eventual microscopic constituents of the
macroscopic devices.

In the experiments just described that relate to
the hypothetical world, quantum mechanics correctly
predicts the correlations between $D$ and $E$
(or other ``emittors'' and ``absorbers'') when suitable
experimental configurations are set up.  In these situations,
the theory can be interpreted in (at least) two
broadly different ways.  In the first one,
the theory is understood as applying to
genuine microscopic objects, emitted by $E$
and detected by $D$.  Perhaps these objects
follow Bohmian-like trajectories, or behave
between $E$ and $D$ in some other
way compatible with quantum mechanics.  In the
other interpretation, there are no microscopic
objects whatsoever going from $E$ to $D$.
There may be something like an action at a distance.
At any rate the theory is in that case
interpreted instrumentally,
for the purpose of quantitatively
accounting for correlations in the stochastic
behavior of $E$ and $D$.

In the hypothetical world we are considering,
I believe that both interpretations
are logically consistent and adequate.
Of course, each investigator can find more satisfaction
in one interpretation than in the other.  The
epistemic view of quantum mechanics corresponds to the
instrumentalist interpretation.  It
simply rejects the existence of
microscopic objects that have no other use
than the one of predicting observed
correlations between macroscopic objects.

In the world in which we live, however, the
situation is crucially different.  The electrons,
neutrons, photons, and other particles that diffract
or interfere are the same that one appeals to in
order to explain the structure of macroscopic objects.
Denying their existence, as is done in the approach
of genuine fortuitousness, dissolves such
explanatory power.  Denying that they
have states, as is done in the epistemic view, leaves
one to explain the state of a macroscopic object
on the basis of entities that have no state.
\section{Interpreting quantum mechanics}
The epistemic and related views therefore fail
to account for the second explanatory role of
quantum mechanics.  To reinforce this conclusion,
it is instructive to investigate what it means
to interpret a theory.

With most physical theories, interpretation is rather
straightforward.  But this should not blind us to the
fact that even very familiar theories can in general
be interpreted in more than one way.  A simple
example is classical mechanics.

Classical mechanics is based on a well-defined mathematical
structure.  This consists of constants $m_i$, functions
$\mathbf{r}_i (t)$, and vector fields $\mathbf{F}_i$
(understood as masses, positions, and forces),
together with the system of second-order differential
equations $\mathbf{F}_i = m_i \mathbf{a}_i$.  A
specific realization of this structure consists in
a system of ten point masses
interacting through the $1/r^2$ gravitational
force.  A hypothesis may then
assert that the solar system corresponds to
this realization, if the sun and nine planets are
considered pointlike and all other objects
neglected.  Predictions made on the basis of
this model correspond rather well with reality.
But obviously the model can be made much more
sophisticated, taking into account for instance
the shape of the sun and planets, the planets'
satellites, interplanetary matter, and so on. 

Now what does the theory have to say about
how a world of interacting masses is really like?
It turns out that such a world can be viewed
in (at least) two empirically equivalent but conceptually
very different ways.  The first one consists in asserting
that the world is made only of small (or extended) masses
that interact by instantaneous action at a distance.
The second way asserts that the masses produce
everywhere in space a gravitational field, which
then locally exerts forces on the masses.  These two ways
constitute two different interpretations of the theory.
Each one expresses a possible way of making the
theory true (assuming empirical adequacy).
Whether the world is such that masses instantaneously
interact at a distance in a vacuum, or a
genuine gravitational field is produced throughout
space, the theory can be held as truly realized.

Similar remarks apply to classical electromagnetism.
The mathematical equations can be interpreted
as referring to charges and currents interacting
locally through the mediation of electric and magnetic
fields.  Alternatively, they can be viewed as referring
to charges and currents only, interacting by
means of (delayed) action at a
distance~\cite{wheeler2}.

In this respect, quantum mechanics seems different
from all other physical theories.  There appears to be
no straightforward way to visualize, so to speak,
the behavior of microscopic objects.  This was vividly pointed
out by Feynman~\cite[p.~129]{feynman} who,
after a discussion of Young's
two-slit experiment with electrons, concluded that
``it is safe to say that no one understands quantum
mechanics.  [...] Nobody knows how it can be like
that.''  But the process of interpreting quantum mechanics
lies precisely in taking up Feynman's challenge.
It is to answer the question, ``How can the world
be for quantum mechanics to be true?''

If we adopt this point of view (known as the
semantic view of theories~\cite{giere,suppe}),
we can understand
the function of Bohmian trajectories or, for that matter, of
other interpretative schemes of quantum mechanics.
Each provides us with one clear way that the microscopic
objects can behave so as to
reproduce the quantum-mechanical rules and,
therefore, the observable behavior of macroscopic
objects.  It is true that, just like the
ether in special relativity, they don't lead to
specific empirical consequences.  But although they
could be dispensed with in the hypothetical world
of Sec.~3, they cannot in the real
world unless, just like the ether was eventually
replaced by the free-standing electromagnetic
field, they are replaced by something that
can account for the structure of macroscopic
objects.

In all physical theories other than quantum mechanics,
there are straightforward and credible answers to the
question raised above, of \,``How can the world be for the
theory to be true?''.  In quantum mechanics there
are a number of answers, for instance Bohmian
trajectories, multiple worlds, modal approaches, etc.
None is straightforward, and none gains universal
credibility.  Should we then adopt the attitude
of the epistemic or related views, which decide
not to answer the question?  I believe that, from
a foundational point of view, this is not tenable.
For how can we believe in a theory, if we are not
prepared to believe in any of the ways it can be
true, or worse, if we do not know any way that it can
be true?
\section{Conclusion}
The epistemic view of quantum mechanics is an
attempt to solve or attenuate the foundational
problems of the theory.  We have seen that it would
succeed if quantum mechanics were used only to explain
nonclassical correlations between macroscopic
objects.  But it is also used to explain the
microscopic structure of macroscopic objects.
Interpreting the theory means finding ways that it
can be intelligible.  A number of proposals go a long
way towards this, but much work remains to be done to make
some of them sufficiently clear and precise.
\section*{Acknowledgment}
This work was supported by the Natural Sciences
and Engineering Research Council of Canada.

\begin{thebibliography}{99}
%
\bibitem{einstein} A.~Einstein, B.~Podolsky, and N.~Rosen,
``Can quantum-mechanical description of physical
reality be considered complete?''
\textit{Physical Review},
47: 777--780, May 1935.
%
\bibitem{bennett} C. H. Bennett and G. Brassard,
``Quantum cryptography: public key distribution and coin tossing,''
\textit{Proceedings of the IEEE International Conference
on Computers, Systems and Signal Processing},
New York: IEEE, 1984, pp.~175--179.
%
\bibitem{shor} P. W. Shor,
``Algorithms for quantum computation: discrete
logarithms and factoring,''
\textit{Proceedings of the 35$^{\, th}$ Annual Symposium on
Foundations of Computer Science},
S.~Goldwasser, Ed.
Los Alamitos, CA: IEEE, 1994, pp.~124--134. 
%
\bibitem{hooft} G. 't Hooft,
``Quantum gravity as a dissipative deterministic system,''
\textit{Classical and Quantum Gravity},
16: 3263--3279, October 1999.
%
\bibitem{leggett} A. J. Leggett,
``Testing the limits of quantum mechanics:
motivation, state of play, prospects,''
\textit{Journal of Physics: Condensed Matter},
14: R415--R451, April 2002.
%
\bibitem{rovelli} C. Rovelli,
``Relational quantum mechanics,''
\textit{International Journal of Theoretical Physics},
35: 1637--1678, August 1996.
%
\bibitem{fuchs1} C. A. Fuchs and A. Peres,
``Quantum theory needs no `interpretation',''
\textit{Physics Today},
53: 70--71, March 2000.
%
\bibitem{fuchs2}C. A. Fuchs,
``Quantum mechanics as quantum information
(and only a little more),'' in
\textit{Quantum Theory: Reconsideration of Foundations},
A.~Khrennikov, Ed.
V\"{a}xj\"{o}:  V\"{a}xj\"{o} U. Press, 2002, pp.~463--543. 
Also available as quant-ph/0205039.
%
\bibitem{vermaas} P.~E.~Vermaas,
\textit{A Philosopher's Understanding of Quantum
Mechanics.  Possibilities and Impossibilities
of a Modal Interpretation},
Cambridge: Cambridge U. Press, 1999.
%
\bibitem{everett} H.~Everett III,
``\, `Relative state' formulation of quantum mechanics,''
\textit{Reviews of Modern Physics},
29: 454--462, July 1957.
%
\bibitem{bohm} D.~Bohm,
``A suggested interpretation of the quantum
theory in terms of `hidden' variables (I and II),''
\textit{Physical Review},
85: 166--193, January 1952.
%
\bibitem{marchildon1} L.~Marchildon,
``Why should we interpret quantum mechanics?''
\textit{Foundations of Physics},
34: 1453--1466, October 2004.
%
\bibitem{marchildon2} L. Marchildon,
``Bohmian trajectories and the ether: Where
does the analogy fail?''
\textit{Studies in History and Philosophy of
Modern Physics},
37: 263--274, June 2006.
%
\bibitem{marchildon3} L. Marchildon,
``The epistemic view of quantum states and the ether,''
\textit{Canadian Journal of Physics},
84: 523--529, January 2006.
%
\bibitem{ballentine} L.~E.~Ballentine,
``The statistical interpretation of quantum mechanics,''
\textit{Reviews of Modern Physics},
42: 358-381, October 1970.
%
\bibitem{heisenberg} W.~Heisenberg,
\textit{Physics and Philosophy.  The Revolution
in Modern Science},
New York: Harper, 1958.
%
\bibitem{neumann} J.~von Neumann,
\textit{Mathematical Foundations of Quantum Mechanics},
Princeton: Princeton U. Press, 1955.
%
\bibitem{peierls} R.~Peierls,
``In defence of `measurement',''
\textit{Physics World},
4: 19--20, January 1991.
%
\bibitem{bloch} I.~Bloch,
``Some relativistic oddities in the quantum theory
of observation,''
\textit{Physical Review},
156: 1377--1384, April 1967.
%
\bibitem{ulfbeck} O.~Ulfbeck and A.~Bohr,
``Genuine fortuitousness.  Where did that click come from?''
\textit{Foundations of Physics},
31: 757--774, May 2001.
%
\bibitem{bohr} A.~Bohr, B.~R.~Mottelson, and O.~ Ulfbeck,
``The principle underlying quantum mechanics,''
\textit{Foundations of Physics},
34: 405--417, March 2004.
%
\bibitem{philippidis} C.~Philippidis, C.~Dewdney, and B.~J.~Hiley,
``Quantum interference and the quantum potential,''
\textit{Il Nuovo Cimento}, 52B: pp.~15--28, July 1979.
%
\bibitem{bub1} J.~Bub,
``Why the quantum?''
\textit{Studies in History and Philosophy of Modern Physics},
35: 241--266, June 2004.
%
\bibitem{bub2} J.~Bub,
``Quantum mechanics is about quantum information,''
\textit{Foundations of Physics},
35: 541--560, April 2005.
%
\bibitem{wheeler2} J.~A.~Wheeler and R.~P.~Feynman,
``Classical electrodynamics in terms of direct
interparticle action,''
\textit{Reviews of Modern Physics},
21: 425--433, July 1949.
%
\bibitem{feynman} R.~P.~Feynman,
\textit{The Character of Physical Law},
Cambridge, MA: MIT Press, 1967.
%
\bibitem{giere} R.~N.~Giere,
\textit{Explaining Science.  A Cognitive Approach},
Chicago: U. of Chicago Press, 1988.
%
\bibitem{suppe} F.~Suppe,
\textit{The Semantic Conception of Theories and
Scientific Realism},
Urbana: U. of Illinois Press, 1989.
%
\end{thebibliography}
\end{document}